\documentclass[doublecol,linenumbers]{epl2} 

\usepackage{graphicx} 
\usepackage[usenames,dvipsnames]{xcolor}
\usepackage{amsmath}
\usepackage{amssymb}
\usepackage{color}
\usepackage{psfrag}
\usepackage{wasysym}
\usepackage{float}
\usepackage{epstopdf}
\usepackage[normalem]{ulem}

\newcommand \beq{\begin{equation}}
\newcommand \eeq{\end{equation}}

\newcommand{\Li}{L_I}
\newcommand{\Lo}{L_O}

\newcommand{\lp}{\ell_p}

\newcommand{\pd}{\partial}

\newcommand{\wo}{\delta}
\newcommand{\woc}{\delta_c}
\newcommand{\tdelta}{\tilde{\wo}}
\newcommand{\sqq}{\sigma_{\theta\theta}}
\newcommand{\srr}{\sigma_{rr}}
\newcommand{\Ugas}{{\cal U}_{\mathrm{gas}}}
\newcommand{\Uelast}{{\cal U}_{\mathrm{strain}}}
\newcommand{\Ubend}{{\cal U}_{\mathrm{bend}}}

\newcommand{\kspring}{k}
\newcommand{\ksubs}{\kappa_w}

\newcommand{\bzeta}{\bar{\zeta}}

\title{Wrinkling reveals a new isometry of pressurized elastic shells}
\shorttitle{Wrinkly isometry of shells}
\author{D. Vella\inst{1}\thanks{E-mail:\email{dominic.vella@maths.ox.ac.uk}} \and H. Ebrahimi\inst{2} \and A. Vaziri\inst{2} \and B. Davidovitch\inst{3}}
\shortauthor{D. Vella \etal}

\institute{                    
  \inst{1}  Mathematical Institute, University of Oxford, Woodstock Rd, Oxford, OX2 6GG, UK\\
  \inst{2} Department of Mechanical and Industrial Engineering, Northeastern University, Boston, Massachusetts 02115, USA\\
  \inst{3} Physics Department, University of Massachusetts, Amherst, Massachusetts 01003, USA}


\abstract{
We consider the point indentation of a pressurized, spherical elastic shell.  Previously it was shown that such shells wrinkle once the indentation reaches a threshold value. Here, we study the behaviour of this system  beyond the onset of instability. We show that rather than simply approaching the classical `mirror-buckled' shape, the wrinkled shell approaches a new, universal shape that reflects a nontrivial type of isometry. For a given indentation depth, this ``asymptotic isometry", which is only made possible by wrinkling, is reached in the doubly asymptotic limit of  weak pressure and vanishing shell thickness.
}

\begin{document}

\maketitle

\section{Introduction}

Just as one pokes an object to see how stiff it is, it is common to measure the mechanical properties of small scale objects by indentation, often using an AFM \cite{vaziri06}. In biology and medicine many 
objects of interest are relatively thin capsules 
swollen by osmotic pressure, \emph{e.g.}~viruses \cite{hernando12}, bacteria \cite{arnoldi00}, yeast cells \cite{arfsten10} and the polymerosomes used in drug delivery \cite{gordon04}. Recent
studies have 
focussed on modelling the indentation of pressurized elastic capsules \cite{arnoldi00,gordon04,hernando12,vella12} using numerous 
approaches including finite element methods and analytical techniques.

\begin{figure}
\centering
\psfrag{d}{$\delta$}
\psfrag{c}{\tiny $\sqrt{\delta R}$}
\psfrag{f}{$F$}
\psfrag{p}{$p$}
\psfrag{r}{$r$}
\psfrag{s}{$R$}
\psfrag{t}{$t$}
\psfrag{z}{$\zeta(r)$}
\includegraphics[width=0.85\columnwidth]{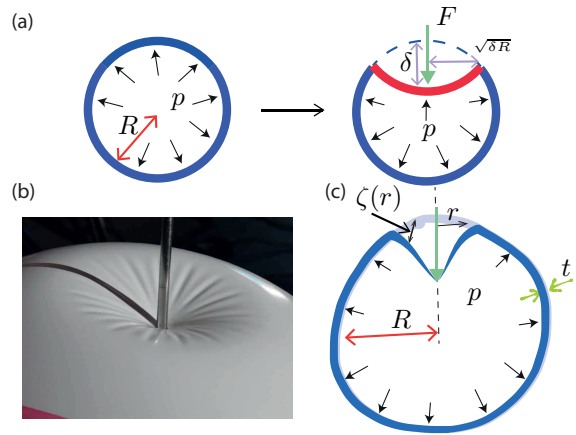}
\caption{(a) Schematic illustration of the mirror buckling of a pressurized shell: an indentation  depth $\delta$ is accommodated by the (energetically cheap) formation of an inverted cap of width $(\delta R)^{1/2}$. (b) In reality, an indented pressurized shell wrinkles \cite{vella11}. (c) Setup and notation for the model calculation of point indention of a pressurized spherical shell presented here.} 
\label{fig:setup}
\end{figure}

The problem of the indentation of a shell with no internal pressure (pressureless) 
has been extensively studied \cite{reissner47a,reissner47b,pogorelov,audoly10}. In this case, the classical picture is that a spherical shell of thickness $t$ and radius $R$ will accommodate indentation by `mirror-buckling': it is as if a portion of the shell were sliced off, inverted and then reattached to the remainder of the shell (fig.~\ref{fig:setup}a). In reality, the apparent discontinuity in slope is resolved by a boundary layer of width $(tR)^{1/2}$ \cite{pogorelov}. Since mirror-buckling represents a simple isometry of the shell (i.e.~there is no strain except at isolated points or along curves), the energy of deformation is localized within this boundary layer and vanishes in the limit $t\to 0$. 
This simple geometrical picture 
has recently been shown to hold even in the presence of an internal pressure, $p$, albeit under the assumption that the shell remains very close to an axisymmetric shape \cite{vella12}.

However, it is well known that deformed shells often do not remain perfectly axisymmetric. Under point indentation, pressureless shells develop 
vertices \cite{fitch68,vaziri08,vaziri09,nasto13,nasto14} while pressurized shells wrinkle (fig.~\ref{fig:setup}b) \cite{vella11}. A similar phenomenology is observed in other deformation modes including the contact between a plate and a shell \cite{audoly10,ebrahimi14} and the secondary buckling of depressurized shells \cite{knoche14a,knoche14b}.

 As was shown  in previous work \cite{vella11}, the appearance of wrinkles upon the point indentation of a pressurized shell is indicative of azimuthal compression. 
Since the mirror-buckled shape remains an isometry of the shell, it is tempting to assume that the wrinkled shape will be a small perturbation of this shape, as has been proposed previously \cite{vella11}. However, here we show that wrinkling changes the shape of the shell markedly. Our analysis reveals a new wrinkly isometry of the shell that is energetically cheaper than mirror buckling. 

\section{Numerical experiments}

Numerical simulations were performed using the commercial finite element package ABAQUS (SIMULIA, Providence, RI) for a spherical shell of radius $R=1\mathrm{~m}$, thickness in the range $0.3\mathrm{~mm}\leq t\leq 2\mathrm{~mm}$,  Young's modulus $E=70\mathrm{~GPa}$, and Poisson ratio $\nu=0.3$. The shell is subject to an internal pressure $p=O(10^5\mathrm{~Pa})$. (Four-node thin shell elements with reduced integration were used in all calculations and a mesh sensitivity study was carried out to ensure that the results are minimally sensitive to the element size. No initial geometrical imperfection was used and the ABAQUS automatic stabilization scheme, with a dissipated energy fraction of $2\times10^{-4}$ and an adaptive stabilization ratio of $0.05$, was employed to capture instabilities in the shell.)

Simulations show that for sufficiently small indentation depths $\wo$, the shell remains perfectly axisymmetric. In this limit the (thickness integrated) in-plane stress \cite{audoly10} is close to the isotropic tension prior to indentation: $\sigma_\infty = pR/2 $. Deformation of the shell is resisted by a normal force that is proportional to the vertical displacement --- effectively a  Winkler foundation of stiffness $\ksubs$ \cite{vella12}.  This foundation stiffness depends on the stretching stiffness of the shell, $Et$, and so on dimensional grounds we might expect that $\ksubs\sim Et/R^2$; an analysis of the governing equations confirms this expectation \cite{reissner47a,vella12}. By analogy with a liquid's meniscus, the  combination of an intrinsic tension, $\sigma_\infty \sim pR$, and a linear foundation stiffness, $\ksubs\sim Et/R^2$,  gives rise to a characteristic horizontal length over which vertical deflections decay  
\beq
\lp=\left(\frac{\sigma_\infty}{\ksubs}\right)^{1/2}=\left(\frac{pR^3}{Et}\right)^{1/2}.
\eeq  The length $\lp$ is an effective capillary length.

Another restoring force is due to the bending rigidity, $B=Et^3/[12(1-\nu^2)]$. 
However, previous work \cite{vella11,vella12} has shown that the effect of bending may be neglected when the internal pressure, $p$, is sufficiently large. To see this, we note that the foundation stiffness resulting from bending effects $\sim B/\lp^4$, which is significantly smaller than the foundation stiffness $\ksubs\sim Et/R^2$ (discussed above) when 
\beq
\tau=\frac{pR^2}{(EtB)^{1/2}}\gg1.
\eeq The parameter $\tau$ may be thought of as the dimensionless pressure within the shell. Since $\tau$ is the ratio between the tension--induced and bending--induced stiffnesses of the shell, it is also similar to the ``tensional bendability" 
introduced in recent studies of uniaxially stretched sheets \cite{davidovitch11}. We  neglect the effects of bending since  all simulations reported here have $\tau\gtrsim 25$. In the language of \cite{hohlfeld15}, we consider the limit of large bendability.

\begin{figure}
\centering
\psfrag{d}[c]{$\tdelta=\wo R/\lp^2$}
\psfrag{L}[c]{$\Li,\Lo/\lp$}
\includegraphics[width=0.75\columnwidth]{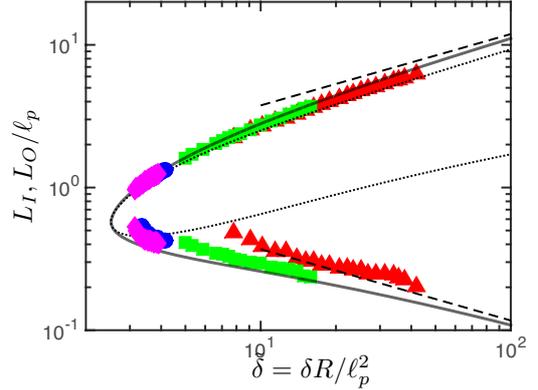}
\vspace{0mm}
\caption{The evolution of the inner and outer wrinkle positions, $\Li$ and $\Lo$, with increasing indentation depth $\tdelta$. The results of ABAQUS simulations are shown for different values of the dimensionless pressure $\tau$ and thickness ratio $t/R$: $\tau=26.2,t/R=3\times10^{-4}$ (red triangles), $\tau=52.5,t/R=3\times10^{-4}$ (green squares), $\tau=177,t/R=2\times10^{-3}$ (blue circles) and $\tau=378,t/R=1\times10^{-3}$ (magenta diamonds). The predictions of the FT membrane theory (valid as $\tau\to\infty$) are shown (solid black curve) together with the asymptotic results  \eqref{eqn:FTout} and \eqref{eqn:FTin} (dashed lines). For comparison, the results of the axisymmetric membrane-shell theory \cite{vella11} are also shown (dotted curve).}
\label{fig:wrinkpos}
\end{figure}

The picture outlined above holds while the stress within the shell remains close to its pre-indentation level. However, indentation also introduces a geometrical strain $\epsilon\sim(\wo/\lp)^2$,  and a corresponding contribution to the stress  
$\sim Et\wo^2/\lp^2$. This additional stress  leads to an inhomogeneity in the stress within the shell: the radial stress  $\srr(r)$ decreases monotonically to the far-field value, $\sigma_\infty=pR/2$, while the hoop stress $\sqq$ decreases more rapidly and overshoots the far-field value (since material circles are pulled inwards and become relatively compressed) \cite{vella11,vella15}. The relative size of the geometrical stress $\sim Et\wo^2/\lp^2$ and the pre-stress $\sim pR$ is a function only of the ratio: 
\beq
\tdelta=\wo R/\lp^2  \ = \ Et\wo  / p R^2.
\label{eqn:woNDdefn}
\eeq Here, this dimensionless indentation depth plays  the role of the ``confinement" parameter introduced in previous studies \cite{king12,grason13,vella15}. Above a critical value,  $\tdelta>\tdelta_c \approx 2.52$, the hoop stress becomes compressive, $\sqq<0$, beginning a  distance $r_c\approx0.58\lp$ from the indenter and confined to an annulus  $\Li<r<\Lo$ as $\tdelta$ increases \cite{vella11}. Since  we  assume that $\tau \gg 1$, (the shell has negligible resistance to compression), any compressive stress  immediately leads to wrinkling. Assuming that the shell remains close to its axisymmetric state,  detailed calculations \cite{vella11}  show that as $\tdelta$ increases, wrinkles are confined to an annulus $\Li<r<\Lo$ with both $\Li$ and $\Lo$  increasing (see fig.~\ref{fig:wrinkpos}).

Simulations confirm the behaviour at the onset of wrinkling\cite{vella11}, in particular the prediction $\tdelta_c\approx2.52$. However, as fig.~\ref{fig:wrinkpos} shows, systematic discrepancies are observed in the wrinkle position as indentation increases beyond $\tdelta_c$. In this letter, we show that this discrepancy signals an important change in the mechanical response.

\section{Tension field theory description of wrinkling}

The key ingredient missing from previous works \cite{vella11,vella12} is that the hoop stress, $\sqq$, cannot become arbitrarily large and negative. Instead, the appearance of wrinkles allows the shell to relax the compressive stress so that $\sqq\approx0$, a result that is referred to variously as Tension field theory \cite{stein61,steigmann90}, Relaxed energy \cite{pipkin86,bella14} or Far from threshold (FT) wrinkling \cite{davidovitch11,davidovitch12}.

An immediate consequence of the collapse of hoop stress, $\sqq\approx0$,  
is that the radial stress $\srr\approx C/r$ in the wrinkled region ($\Li<r<\Lo$), for some constant $C$. We denote the normal displacement of the wrinkled shell from its natural state  by $\zeta(r,\theta)=\bzeta(r)+f(r)\cos m\theta$: $\bzeta$ is the mean displacement upon which $m$ fine wrinkles of amplitude $f(r)$ are placed. (Here  $r$ is the horizontal distance from the axis of indentation and $\theta$ the azimuthal coordinate, see fig.~\ref{fig:setup}c.) In the FT approach, $\bzeta(r)$ is determined at leading order with $m$ and $f$ determined subsequently \cite{davidovitch11,king12}. In the wrinkled zone  vertical force balance reads \cite{vella15b}
\beq
\frac{C}{r}\left(\frac{1}{R}-\frac{\upd^2\bzeta}{\upd r^2}\right)=p-\frac{F}{2\pi}\frac{\delta(r)}{r},
\eeq which can immediately be integrated to give
\beq
\bzeta=A+\frac{r^2}{2R}-\frac{p r^3}{6C}+\frac{F}{2\pi C}r
\label{eqn:FTzeta}
\eeq for some constant of integration $A$ and the force $F$.

To determine $A$, $F$, and $C$, together with the inner and outer limits of the wrinkled region, $\Li$ and $\Lo$, the above solution must be patched together with the solution of the vertical force balance in the tensile (i.e.~unwrinkled) regions $0\leq r<\Li$ and $r>\Lo$, which reads \cite{vella11,vella15b}
\beq
\frac{1}{r}\frac{\upd}{\upd r}\left[\left(\frac{r}{R}-\frac{\upd \zeta}{\upd r}\right)\psi\right]=p-\frac{F}{2\pi}\frac{\delta(r)}{r}.
\label{eqn:TensFvK1}
\eeq Here $\psi(r)$ is the Airy potential, such that $\srr=\psi/r$ and $\sqq=\psi'$, which satisfies the compatibility condition \cite{vella11}
\beq
\frac{1}{Y}r\frac{\upd}{\upd r}\left[\frac{1}{r}\frac{\upd}{\upd r}\left(r\psi\right)\right]=\frac{r}{R}\frac{\upd \zeta}{\upd r}-\tfrac{1}{2}\left(\frac{\upd \zeta}{\upd r}\right)^2.
\label{eqn:TensFvK2}
\eeq

Each of the equations in the tensile regions, \eqref{eqn:TensFvK1} and \eqref{eqn:TensFvK2}, is second order; there are therefore 4 degrees of freedom associated with each tensile region. In the wrinkled region there are 3 degrees of freedom ($A$, $C$ and $F$). Finally, with the radii of the wrinkled zone $\Li,\Lo$, we have  
13 degrees of freedom in total, which must match the number of boundary conditions (BCs). At the indenter, $r=0$, we impose the vertical displacement, $\zeta(0)=-\delta$, while the radial displacement vanishes, $u_r(0)=0$. Far from the indenter, we require that $\zeta(r)\to0$, while $\psi\approx pRr/2$, the uniformly stressed solution. In addition, we also require continuity of $\zeta$, $\zeta'$, $\srr$, $\sqq$ and $u_r$,  at $r=\Li$ and $r=\Lo$. The first four of these continuity conditions at each boundary, together with the conditions at $0$ and $\infty$, give a total of $12$ BCs. Imposing continuity of $u_r$ at $\Li$ and $\Lo$ requires the integration of a further equation for $\pd u_r/\pd r$ across the wrinkled region, introducing a new unknown and two BCs. After  some algebra a single condition emerges
\beq
\frac{C}{Y}\log\frac{\Lo}{\Li}=\int_{\Li}^{\Lo}\tfrac{1}{2}\left(\frac{\upd\bzeta}{\upd r}\right)^2-\frac{r}{R}\frac{\upd\bzeta}{\upd r}~\upd r.
\label{eqn:FTclosure}
\eeq Here, the LHS and RHS represent, respectively, the stretching of lines of longitude, and how the boundary displacement differs from that expected were the shell to deform inextensibly. 


Equations \eqref{eqn:FTzeta}--\eqref{eqn:FTclosure} can be solved numerically using the MATLAB multipoint boundary value problem solver \texttt{bvp4c}. For a given $\wo>\woc$, the solution gives the inner and outer limits of the wrinkled region ($\Li$ and $\Lo$ respectively), together with the force $F$. Figure \ref{fig:wrinkpos} shows the evolution of $\Li$ and $\Lo$ with $\tdelta$, together with the results of our ABAQUS simulations. Also shown for comparison are the extent of the compressive hoop stress for a perfectly axisymmetric membrane-shell \cite{vella11}. The agreement between simulations and the results of this collapsed-compressive stress model, which takes into account the non-perturbative effect of wrinkling on the shell shape, is significantly better than that based on the axisymmetric membrane-shell theory. In particular, we observe that $\Li$ decreases with increasing $\tdelta$ --- qualitatively different from the predictions of a standard post-buckling analysis \cite{vella11}. 
%

\begin{figure}
\centering
\psfrag{d}{$\tdelta$}
\psfrag{e}{\tiny $\tdelta$}
\psfrag{F}[c]{$F/p\lp^2$}
\psfrag{k}[c]{\tiny $\kspring=F/(pR\wo)$}
\includegraphics[width=0.75\columnwidth]{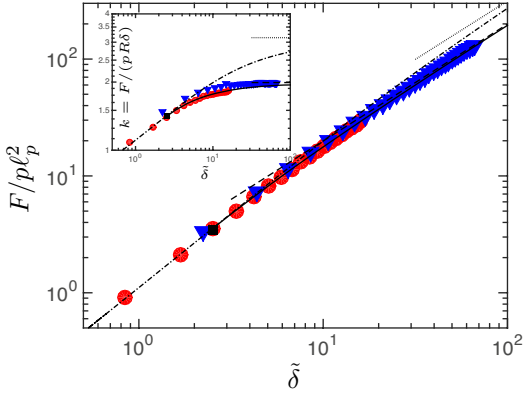}
\vspace{0mm}
\caption{Numerical results for the force--displacement relationship of an indented pressurized shell. The results of ABAQUS simulations are shown for a shell thickness $t/R=3\times10^{-4}$. Different pressurisations are used: $p=0.1\mathrm{~MPa}$, $\tau\approx52.5$ (red circles) and $p=0.05\mathrm{~MPa}$, $\tau\approx26.2$ (blue triangles). The  numerical solutions of \eqref{eqn:FTzeta}--\eqref{eqn:FTclosure} are shown (solid black curve) with the asymptotic relationship \eqref{eqn:FTforcelaw} (dashed black line). The corresponding results for a perfectly axisymmetric shell are also shown (dash-dotted curve) with the ultimate mirror-buckling result $F=\pi \delta pR$ (dotted line)  \cite{vella12}. The black square shows the onset of wrinkling, at $\tdelta\approx2.52$.}
\label{fig:forcelaw}
\end{figure}

Figure \ref{fig:forcelaw} shows the force--displacement relationship: ABAQUS simulations and the numerical solution of \eqref{eqn:FTzeta}--\eqref{eqn:FTclosure} produce results that are in good agreement with each other. Furthermore, in the limit of large indentation depths, the force-displacement relationship becomes linear, corresponding to a constant stiffness $k=F/pR\wo$ (inset of fig.~\ref{fig:forcelaw}). At the scaling level, this result is in qualitative agreement with prior analysis of the axisymmetric membrane-shell equations and with the simple picture of deformation being accommodated via mirror-buckling \cite{vella12}: the work done $\sim F\delta$ is  stored purely in the gas compressed at constant pressure $p\Delta V\sim p[(\wo R)^{1/2}]^2\wo$, so that $F\sim pR\delta$. However, for the axisymmetric, mirror-buckled shape, a detailed calculation predicts that $k=F/pR\wo=\pi$ \cite{vella12} while fig.~\ref{fig:forcelaw} indicates that $k \approx 2$. To gain quantitative understanding of the differences between the axisymmetric membrane theory and  the compressed-stress approach described so far, we now consider the limit of very large indentation depths, $\tdelta\gg1$.

\section{Analytical results for large indentation}

As already remarked, $\Li\to0$ as $\wo$ increases. This can be shown by a more formal calculation \cite{vella15b} but for now we make this an ansatz that we verify \emph{a posteriori}. With this assumption, it is clear that the constant $A$ in  \eqref{eqn:FTzeta} is $A\approx-\wo=\bzeta(0)$, the imposed indentation depth. A detailed calculation \cite{vella15b} shows that outside the wrinkled region, $r>\Lo$, the vertical deflection of the shell decays exponentially. For the wrinkled zone to patch  onto this we must have  $\bzeta(\Lo)=\bzeta'(\Lo)=0$. The deflection of the wrinkled portion of the shell may then be rewritten
\beq
\frac{\bzeta}{\wo}=-1+\left(\frac{3}{2}-\frac{\lambda^2}{4}\right)\xi+\frac{\lambda^2}{2}\xi^2-\left(2+\lambda^2\right)\xi^3/4
\label{eqn:ndshape}
\eeq where $\xi=r/\Lo$ and $\lambda=\Lo/(\wo R)^{1/2}$. Substituting \eqref{eqn:ndshape} into the constraint \eqref{eqn:FTclosure}, and neglecting the logarithmic term, we find $\lambda=[6(\sqrt{5}-2)]^{1/2}\approx1.19$ so that  for $\tdelta\gg1$:
\beq
\Lo\approx 1.19013(\wo R)^{1/2}
\label{eqn:FTout}
\eeq

Comparing coefficients between  Eqs.~\eqref{eqn:ndshape} and \eqref{eqn:FTzeta} we find:
\beq
C=\frac{\lambda^3}{3(1+\lambda^2/2)}p\wo^{1/2}R^{3/2}\approx0.329p\wo^{1/2}R^{3/2} \ , 
\label{eqn:FT_C}
\eeq 
\beq
F\sim6\pi C\wo\frac{3-\sqrt{5}}{2\lambda (\wo R)^{1/2}}\approx1.99pR\,\wo.
\label{eqn:FTforcelaw}
\eeq

To determine how $\Li$ vanishes with increasing indentation depth we use a scaling argument. Recall that the radial stress in the wrinkled region is given by $\srr=C/r$. The stress in the inner tensile region ($r<\Li$) may be estimated by using the geometrically induced strain in the deformed region, $r<\Lo$, giving $\srr\sim Et (\wo/\Lo)^2$ (a similar estimate for the stress in a tensile inner core  has been used in a related problem \cite{vella15}). Equating these two stresses at  the inner edge of the wrinkles, we find that $\Li  \sim  p R^{5/2}\wo^{-1/2}/(Et)$.  A detailed calculation \cite{vella15b} yields:
\beq
\frac{\Li}{\lp}  \approx 1.16923 \left(\frac{\lp^2}{\wo R}\right)^{1/2}.
\label{eqn:FTin}
\eeq

\noindent The predictions \eqref{eqn:FTout}, \eqref{eqn:FTforcelaw} and \eqref{eqn:FTin} are in excellent agreement with the solution of the full tension-field problem and with ABAQUS simulations (figs~\ref{fig:wrinkpos} and \ref{fig:forcelaw}).

We note that the force law in \eqref{eqn:FTforcelaw} has the same scaling behaviour as the perfectly axisymmetric case \cite{vella12}, but with prefactor $1.99$ rather than $\pi$. In particular, this scaling is independent of the elastic properties of the shell ($E$ and $t$) depending only on the internal pressure and shell radius. In the perfectly axisymmetric shell, this independence of material properties was a natural consequence of mirror-buckling being an isometry of the shell \cite{vella12}.  We shall see that an asymptotic isometry of the wrinkled shell (i.e.~with the compression collapsing faster than the residual tension), underlies the form of the force law in \eqref{eqn:FTforcelaw}.

\begin{figure}
\centering
\psfrag{r}[c]{$r/(\wo R)^{1/2}$}
\psfrag{y}[c]{$(y-R)/\wo$}
\includegraphics[width=0.75\columnwidth]{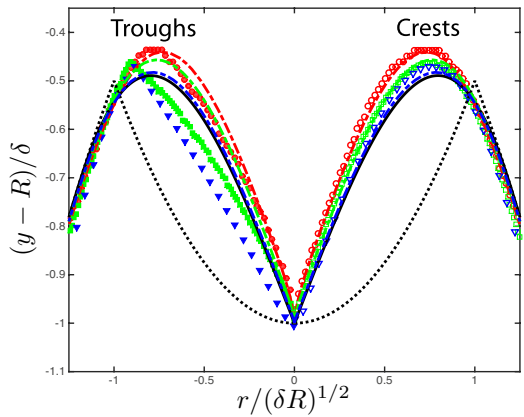}
\vspace{0mm}
\caption{Approach to the wrinkly isometry of a pressurized shell. Cross-sections through an indented pressurized shell, obtained from ABAQUS simulations, are shown for different indentation depths  with $\tau\approx52.5$: $\tdelta=6.1$ (red circles) and $\tdelta=15$ (green squares); to reach larger indentations numerically we must consider lower pressure, $\tau=26.2$ and $\tdelta\approx90.3$ (blue triangles). Cross-sections through a trough in the wrinkle pattern are shown on the left ($r<0$); Cross-sections through a crest are shown on the right ($r>0$). The predictions of the FT theory with the corresponding (finite) value of $\tdelta$ are shown by the coloured dash-dotted curves. As $\tdelta$ increases the deformation tends to the wrinkly isometry \eqref{eqn:UniversalShape} (solid black curve), rather than the classic mirror-buckled isometry (dotted curve).}
\label{fig:xsection}
\end{figure}

Finally, using the various pre-factors determined by the preceding analysis, we rewite Eq.~\eqref{eqn:ndshape}:
%
\beq
\frac{\bzeta}{\wo}\approx-1+0.963\frac{r}{(\wo R)^{1/2}}+\frac{1}{2}\frac{r^2}{\wo R}-0.507\frac{r^3}{(\wo R)^{3/2}},
\label{eqn:UniversalShape}
\eeq 
for $1.17\lp^2/(\wo R)^{1/2}\lesssim r\lesssim1.19(\wo R)^{1/2}$. When written in this way the profile in \eqref{eqn:UniversalShape} appears universal. Fig.~\ref{fig:xsection} shows that cross-sections of simulated shells 
tend to this universal shape as the confinement $\tdelta$ increases. In particular, we see that \eqref{eqn:UniversalShape} gives a much better account of numerical simulations than classic mirror-buckling, which, in this notation would read $\zeta/\wo\approx-1+r^2/(\wo R)$. 
A puzzling observation is the asymmetry of wrinkle crests and troughs: the profile \eqref{eqn:UniversalShape} captures  the former better than the latter.


\section{The energy stored in the indented shell}
Let us evaluate now the elastic energy stored in the shell at large indentation depth, when $\Li/\lp \to 0$. This energy consists of two principal parts: the stretching energy $\Uelast$ stored in the tensile component $\sigma_{rr} = C/r$ of the compression-free stress field, and the bending energy $\Ubend$ of the wrinkles. The evaluation of $\Uelast$ is straightforward:        
\beq
\Uelast=\pi\int_{\Li}^{\Lo}~r\sigma_{rr}\epsilon_{rr}~\upd r\sim \frac{p^2R^3}{Et}\wo \log\tdelta \ , 
\label{eqn:FTUelast}
\eeq 
where we used Eq.~(\ref{eqn:FT_C}). A more careful calculation of $\Uelast$, which accounts also for  the tensile core zone $r<\Li$, confirms this scaling relation.  

A calculation of $\Ubend$, similarly to previous works \cite{king12,hohlfeld15}, proceeds in two steps. First, we express the energy as a function of $m$ only, by employing a ``slaving" condition between the number, $m$, and amplitude, $f(r)$, of wrinkles:
\beq
\frac{m^2f^2}{4r}=\frac{C}{Y}\log\frac{\Lo}{r}+\int_r^{\Lo} \!\!\upd r\left[\frac{r}{R}\frac{\upd \bzeta}{\upd r}-\tfrac{1}{2}\left(\frac{\upd \bzeta}{\upd r}\right)^2\right]. 
\label{eqn:FTslaving}
\eeq 
Here the RHS may be calculated from the results of the preceding analysis, Eqs.~(\ref{eqn:FTzeta},\ref{eqn:FTout},\ref{eqn:FT_C},\ref{eqn:FTforcelaw}). 
%
Physically, \eqref{eqn:FTslaving} stems from the fact that, for the compressive stress to collapse, the azimuthal undulations must accommodate the excess length of deformed latitudes. (Notice also that Eq.~(\ref{eqn:FTclosure}) can be obtained from the slaving condition (\ref{eqn:FTslaving}) since the wrinkle amplitude vanishes at $r=\Li,\Lo$.) 

In the second step, one determines the energetically favorable value of $m$, for which the bending energy balances the contribution of wrinkles to the stretching of longitudes (beyond $\Uelast$). Intuitively, bending favors small curvature (i.e.~small $m$) while stretching resists deviations from 
$\bzeta(r)$, thus favoring small $f(r)$ or, by condition (\ref{eqn:FTslaving}), large $m$. The detailed calculation \cite{vella15b} yields:  
\beq
m\sim \left(\frac{\tdelta^2}{\log\tdelta}\right)^{1/4}\tau^{1/2} \ , 
\label{eqn:scaling-m}
\eeq 
and gives the bending energy
\beq
\Ubend \sim pR\wo^2(\log\tdelta)^{1/2}/\tau . 
\label{eqn:FTUbend}
\eeq
As shown above, our ABAQUS simulations are in excellent agreement with the macroscale features of the wrinkle pattern predicted theoretically ($\Li$, $\Lo$, $F$ and the shell profile). However, while close to threshold (i.e.~$\tdelta=O(1)$), previous ABAQUS simulations \cite{vella11} confirm the $\tau$ dependence predicted by (\ref{eqn:scaling-m}), the predicted evolution of $m$ with increasing $\tdelta$ has not been observed. This shortcoming is consistent with recent work, which showed significant hurdles in numerically predicting the wrinkle number \cite{Taylor15}. In that study, which addressed a planar sheet under differential tensile loads, it was shown that under-damped dynamic relaxation allows ``hopping" between meta-stable states, leading ultimately to the ground state (i.e.~the wrinkle number of the global energy  minimizer). In our simulations, the lack of variation of wrinkle number persists despite our use of an over-damped relaxation technique, suggesting either a sensitivity to numerical technique or a higher energetic barrier between meta-stable states than in other problems \cite{Taylor15}. Additionally, a more elaborate calculation of the energy may result in a spatial profile of the wrinkle number, $m \to m(r)$ \cite{pineirua13,bella14, InPrep15}.

\section{Wrinkly isometry}

A key feature of \eqref{eqn:UniversalShape} in the limit of very large indentations, $\tdelta\gg1$, is that the wrinkle pattern covers almost the entire 
region from the indenter to the outer ridge, which is positioned at $r\approx1.19(\wo R)^{1/2}$; outside this ridge, the shell is effectively undeformed \cite{vella15b}. As such, the work done in compressing the gas is: 
\beq
\Ugas=p\Delta V=2\pi p\int_{\Li}^{\Lo}r\bzeta~\upd r\approx0.995p R\wo^2.
\label{eqn:FTUgas}
\eeq 
Interestingly, the derivative $\upd\Ugas/\upd\wo$ is  identical to the total force $F$  calculated above (Eq.~\ref{eqn:FTforcelaw}). The origin of this remarkable feature is  elucidated by comparing $\Ugas$ to the energies of stretching and bending stored in the poked shell. We have that $\Uelast/\Ugas\sim (\log\tdelta)/\tdelta\ll1$, since \eqref{eqn:UniversalShape} only holds for $\tdelta\gg1$. At the same time, for the FT expansion \cite{davidovitch11} to be valid we must have $\Ubend\ll\Ugas$, i.e.~$\tau\gg(\log\tdelta)^{1/2}$. The two components of the elastic energy are therefore negligible in comparison to the work done in compressing the gas at constant pressure; this is the signature of the ``asymptotic isometry" that is realized by the wrinkled shell.

Rather than a specific shape, such as mirror-buckling, which exists also in the absence of pressure, and to which the undeformed shell can be mapped isometrically almost everywhere (except at the circular ridge), a wrinkled shell develops finer and finer structure in a macroscopic zone ($\Li<r<\Lo$). However, this isometric shape can only be approached when both $\Ubend,\Uelast\ll\Ugas$, i.e.~when
\beq
1\ll\tdelta\ll\exp(\tau^2).
\label{eqn:Inequalities-regime} 
\eeq We shall show shortly that this occurs when both the shell thickness and the pressure become asymptotically small. Such a doubly asymptotic approach to isometry has been noted recently in set-ups that involve strong deformation of thin sheets; in these systems the work done by the external load is generally stored elsewhere than the deformed sheet itself \cite{vella15, chopin15, hohlfeld15}.

\section{Conclusion} Why is the wrinkly isometry selected by the poked shell, and not the simpler, mirror-buckled isometry? In the parameter regime (\ref{eqn:Inequalities-regime}) the deformation energy is well-approximated by $\Ugas$ alone: the energies of bending and stretching (which are localized in the circular ridge of the mirror-buckled shape but spread around in the wrinkly isometry) are insignificant. The only energy that one should compare then is the work of compressing the gas. In both types of isometric response this energy $\propto \ p R \delta^2$ but for wrinkly isometry the prefactor $\approx0.995$ while for mirror-buckling it is $\pi/2$: since wrinkly isometry compresses less gas than mirror-buckling, it is, in this limit, energetically favourable to mirror-buckling.


To understand the physical nature of the  asymptotic parameter regime \eqref{eqn:Inequalities-regime}, we note that \eqref{eqn:Inequalities-regime} may be re-written $1\ll \delta/(t\tau)\ll\exp(\tau^2)$. For a fixed indentation depth $\delta$, the first inequality may be achieved if, $t\tau\to0$; however, the second inequality then holds only if $\tau\to\infty$. If the Young's modulus and radius of the shell are considered fixed (as in our simulations) this  may be achieved with vanishing thickness and pressure, $t,p\to0$, but with the additional requirements that  $p/t\to0$ and $p/t^2\to\infty$ (e.g.~$p\sim t^\alpha$ for any $1<\alpha<2$). The wrinkly isometry therefore exists if the pressure is sufficiently weak, but not too weak.

If instead, $p/t^2\to0$ as $t\to0$, the pressure is ``ultra-weak". In this case, the energy of deformation is not well-approximated by $\Ugas$, and the bending energy of deformation must be considered. One might guess that in such a case the deformed shell will approach isometry simply by selecting the mirror buckled shape. A host of experimental and computational studies \cite{vaziri08,vaziri09}, as well as one's own experience of poking a ping-pong ball, indicate that such a guess is naive: instead, a polygonal shape is formed, with a number of vertices that increases slowly upon increasing indentation depth (relative to thickness). Such a shape is strictly different  from both the axisymmetric mirror-buckled shape and the wrinkly isometry described here (for a weak, but not ultra-weak, pressure). This suggests the fascinating possibility that a new type of isometry characterizes the poked shell for small thicknesses and ultra-weak pressure. We hope that future work will clarify the nature of this third type of isometry of shells.     

\begin{acknowledgments}
We thank the Aspen Center for Physics and the Parsegians' Casa F\'{i}sica for hospitality during this work. This research was supported by a Leverhulme Trust Research Fellowship (D.V.), the Zilkha Trust, Lincoln College (D.V.), the European Research Council under the European  Horizon 2020 Programme, ERC Grant Agreement no. 637334 (D.V.), NSF-CAREER Grant No. CMMI-1149750 (H.E and A.V.),  NSF-CAREER Grant No. DMR 11-51780 (B.D.) and  a fellowship from the Simons Foundation, Award No. 305306, (B.D.).
  
\end{acknowledgments}

\end{document}


\maketitle

\section{Introduction}

In this Supplementary Information, we present more details of the calculation presented in the main text. In particular in \S\ref{sec:largeD} we present the details of the calculation of the base shape, $\zeta_0(r)$, about which wrinkles occur. The main results of this analysis (which are also given in the paper) are summarized in \S\ref{sec:largeD}\ref{sec:summary}. In \S\ref{sec:subdom} we estimate the energy of the wrinkle pattern and thereby determine the scaling law for the wrinkle number given in the main paper. Finally, in \S\ref{sec:compare} we compare the results for the indentation of a pressurized shell with those obtained for the indentation of an indented floating sheet \cite[][]{vella15}.

%

\section{Governing equations}

We begin with the dimensionless equations given by \cite{vella11} for the deformation of a pressurized shell. For consistency with the main text we use $\zeta(r)$ to denote the normal displacement of the shell from its pressurized, pre-indented state. We set
\begin{equation}
\lp=R\left(\frac{pR}{Y}\right)^{1/2},\quad \wr=\frac{R\zeta}{\lp^2},\quad \rho=\frac{r}{\lp},\quad \Psi=\frac{\psi}{pR\lp},\quad F=\frac{f}{p\lp^2},
\label{ndim:p>0}
\end{equation} where $Y=Et$ is the stretching stiffness of the shell. With this non-dimensionalization, we find that the shallow shell equations \cite[][]{ventsel01,vella11} become
\beq
\tau^{-2}\nabla^4\wr+\frac{1}{\rho}\frac{\upd }{\upd \rho}\left(\rho\Psi\right)-\frac{1}{\rho}\frac{\upd }{\upd \rho}\left(\Psi\frac{\upd \wr}{\upd \rho}\right)=1-\frac{F}{2\pi}\frac{\delta(\rho)}{\rho}
\label{fvk1:p>0}
\eeq and
\beq
\rho\frac{\upd }{\upd \rho}\left[\frac{1}{\rho}\frac{\upd }{\upd \rho}\left(\rho\Psi\right)\right]=\rho \frac{\upd \wr}{\upd\rho}-\frac{1}{2}\left(\frac{\upd \wr}{\upd \rho}\right)^2,
\label{fvk2:p>0}
\eeq
where
\beq
\tau=\frac{pR^2}{(YB)^{1/2}}
\eeq is a dimensionless measure of the tension within the shell due to inflation alone, $\sigma_\infty=pR/2$.

In the main text, we focus on the membrane limit, which corresponds to $\tau=\infty$. We shall therefore neglect the term proportional to $\nabla^4\wr$ in \eqref{fvk1:p>0}. However, when discussing the scaling behaviour of the number of wrinkles, we shall return to consider the role of this term.

A key control parameter is the indentation depth, which in these dimensionless variables is
\beq
\wo=\delta R/\lp^2.
\eeq This imposed indentation depth is imposed by the boundary condition $\wr(0)=-\wo$, which in turn requires a point force $F$ applied at $\rho=0$. A key quantity of interest is the relationship between $F$ and $\wo$.

\subsection{Tensile regions} 

In regions where both the radial and hoop stresses remain tensile, i.e.~$\srr,\sqq>0$, the shell deformation is described by the membrane-shell equations
\beq
\frac{1}{\rho}\frac{\upd }{\upd \rho}\left(\rho\Psi\right)-\frac{1}{\rho}\frac{\upd }{\upd \rho}\left(\Psi\frac{\upd \wr}{\upd \rho}\right)=1-\frac{F}{2\pi}\frac{\delta(\rho)}{\rho}
\label{fvk1:mem}
\eeq and
\beq
\rho\frac{\upd }{\upd \rho}\left[\frac{1}{\rho}\frac{\upd }{\upd \rho}\left(\rho\Psi\right)\right]=\rho \frac{\upd \wr}{\upd\rho}-\frac{1}{2}\left(\frac{\upd \wr}{\upd \rho}\right)^2.
\label{fvk2:mem}
\eeq 

\subsection{The wrinkled region}

It has been shown previously \cite{vella11} that the hoop stress becomes compressive ($\sqq<0$) for sufficiently large indentation depths, $\wo>\woc\approx2.52$. In reality, a membrane cannot support a compressive stress and so we must instead assume that the compression is completely relaxed, $\sqq\approx0$, within an, as yet unknown, wrinkled region $\Li<\rho<\Lo$. In this region, the equilibrium equation
\beq
\frac{\upd\srr}{\upd r}+\frac{\srr-\sqq}{r}=0
\eeq leads immediately to
\beq
\srr=\frac{C}{r}.
\label{eqn:WrinkledSrr}
\eeq Equation \eqref{eqn:WrinkledSrr} replaces the equation of compatibility of strains \eqref{fvk2:mem} in wrinkled regions.

The appropriate expression of vertical force balance in a wrinkled region may be found  by substituting a dimensionless version of \eqref{eqn:WrinkledSrr}, or $\Psi=C'=C/ pR\lp$, into \eqref{fvk1:mem}. Dropping the primes, we have
\beq
\frac{\upd^2 \wr}{\upd \rho^2}=1-\frac{\rho}{C}+\frac{F}{2\pi C} \delta(\rho)
\label{eqn:WrinkledFB}
\eeq

\subsection{Matching boundary conditions}

Equations \eqref{fvk1:mem}--\eqref{fvk2:mem} were solved in the tensile regions $0<\rho<\Li$ and $\rho>\Lo$ while equation \eqref{eqn:WrinkledFB} was solved for the wrinkled region $\Li<\rho<\Lo$. The complete set of boundary conditions applied were
\begin{align}
\wr(0)&=-\wo,\quad \lim_{\rho\to0}\bigl[\rho\Psi'-\nu\Psi\bigr]=0\\
[\wr]_{\Li^-}^{\Li^+}&=[\wr']_{\Li^-}^{\Li^+}=[u_r]_{\Li^-}^{\Li^+}=0,\quad \Psi(\Li^-)=C,\quad \Psi'(\Li^-)=0\\
[\wr]_{\Lo^-}^{\Lo^+}&=[\wr']_{\Lo^-}^{\Lo^+}=[u_r]_{\Lo^-}^{\Lo^+}=0,\quad \Psi(\Lo^+)=C,\quad \Psi'(\Lo^+)=0\\
\wr&\to0,\quad \Psi\sim \rho/2,\quad \mathrm{as} \quad \rho\to\infty.
\end{align}

\section{Analytical results in the limit of large indentations\label{sec:largeD}}


From the numerics, it appears that $\Li\to0$ as $\wo$ increases. At the same time, we expect the general horizontal scale to be increasing (specifically $\sim\wo^{1/2}$) and so we seek to exploit the fact that $\Li/\wo^{1/2}\ll1$ to simplify the problem within the inner tensile region, $\rho<\Li$; in particular,  we will be able to neglect the effect of the pressure loading within this region.

\subsection{The inner tensile region $0<\rho<\Li$\label{sec:inner}}

The numerical solution of the reduced, FT, problem suggest that $\Li\sim \wo^{-1/2}$ and $C\sim\wo^{1/2}$ for $\wo\gg1$. We therefore let $\Li=\lamI\wo^{-1/2}$, $C=\mu\wo^{1/2}$ and rescale the equations by letting
\beq
\xi=\rho/\Li,\quad \omega(\xi)=\wr/\wo,\quad \chi=\Psi/C=\Psi/\mu\wo^{1/2}.
\eeq We find that
\beq
\frac{1}{\xi}\frac{\upd }{\upd \xi}\left(\xi\chi\right)-\frac{\wo^2}{\lamI^2}\frac{1}{\xi}\frac{\upd }{\upd \xi}\left(\chi\frac{\upd \omega}{\upd \xi}\right)=\frac{\lamI}{\mu\wo}-\frac{F}{2\pi\lamI\mu}\frac{\delta(\xi)}{\xi}.
\label{fvk1:tensile}
\eeq While it may seem that there is nothing that can balance the second term on the LHS of \eqref{fvk1:tensile}, it is important to remember that we do not expect $\upd \omega/\upd\xi$ to be $O(1)$ within the tensile region; rather this quantity will be small, and so the correct balance is between the second term on the LHS and the second term on the RHS. Integrating, we find that
\beq
\chi\frac{\upd \omega}{\upd \xi}=\frac{F\lamI}{2\pi\mu\wo^2}=\cF.
\label{fvk1:tensileapprox}
\eeq This now gives us an expression with which to eliminate $\omega'$ from the equation of compatibility, which in rescaled coordinates reads
\beq
\xi\frac{\upd }{\upd \xi}\left[\frac{1}{\xi}\frac{\upd }{\upd \xi}\left(\xi\chi\right)\right]=\frac{\lamI}{\mu}\xi \frac{\upd \omega}{\upd\xi}-\frac{1}{2}\frac{\wo^2}{\lamI\mu}\left(\frac{\upd \omega}{\upd \xi}\right)^2.
\eeq Assuming that $\omega'>O(\wo^{-2})$, i.e.~that $F\gtrsim O(1)$, the first term on the RHS, which arises from the curvature of the shell, may be neglected in comparison to the second, which arises from the strain caused by out-of-plane deformation. We then find that
\beq
\xi\frac{\upd }{\upd \xi}\left[\frac{1}{\xi}\frac{\upd }{\upd \xi}\left(\xi\chi\right)\right]\approx-\frac{1}{2}\frac{\wo^2}{\lamI\mu}\left(\frac{\upd \omega}{\upd \xi}\right)^2=-\frac{1}{2}\frac{\wo^2\cF^2}{\lamI\mu}\chi^{-2}.
\label{eqn:innertenschiprob}
\eeq Eqn \eqref{eqn:innertenschiprob} is the usual compatibility of strains equation that is found in the FvK equation for a naturally flat sheet: the curvature of the shell does not play a role in the inner tensile region.

Equation \eqref{eqn:innertenschiprob} is to be solved on $0<\xi\leq1$ with boundary conditions
\begin{align}
\srr(\Li)=C/\Li&\implies\chi(1)=1\label{eqn:innerBC1}\\
\sqq(\Li)=0&\implies\chi'(1)=0\label{eqn:innerBC2}
\end{align} and zero horizontal displacement at the origin, i.e.~$\lim_{\rho\to0}\bigl[\rho\Psi'-\nu\Psi\bigr]=0$.

Eqn \eqref{eqn:innertenschiprob} with boundary conditions \eqref{eqn:innerBC1}--\eqref{eqn:innerBC2} can be solved analytically using  steps similar to those outlined in related problems \cite[see][for example]{chopin08,vella15}. We first let:
\beq
f^2=\frac{\wo^2\cF^2}{\lamI\mu},\quad\eta=\xi^2,\quad\Phi=\xi\chi
\eeq so that
\beq
\frac{\upd^2\Phi}{\upd\eta^2}=-\frac{f^2}{8\Phi^2}
\eeq with boundary conditions
\beq
\Phi(0)=0,\quad \Phi'(1)=1/2,\quad \Phi(1)=1.
\eeq

The first integral gives
\beq
\frac{\upd\Phi}{\upd \eta}=\frac{f}{2}\left(\frac{1+A\Phi}{\Phi}\right)^{1/2}
\eeq where $(1+A)^{1/2}=f^{-1}$ using the boundary conditions at $\eta=1$. Integrating again, we have
\beq
\frac{A}{2(1+A)^{1/2}}(1-\eta)=(1+A)^{1/2}-\Phi^{1/2}(1+A\Phi)^{1/2}+A^{-1/2}\left[\sinh^{-1}(A\Phi)^{1/2}-\sinh^{-1}A^{1/2}\right].
\eeq The condition at $\eta=0$ then gives that the constant $A$ must satisfy
\beq
\frac{A}{2(1+A)^{1/2}}=(1+A)^{1/2}-A^{-1/2}\sinh^{-1}A^{1/2},
\eeq which has solution $A\approx -0.696529 $ and hence $f\approx1.81527$. Note that this is precisely the same result as was found for the indentation of a floating, naturally flat, membrane by \cite{vella15}. This similarity occurs because sufficiently close to the origin neither the internal pressure nor the curvature of the shell play a role: the balance is purely that between the indentation force $F$ and the tension within the deformed membrane.

The result $f\approx1.81527$ shows that ${\cal F}\sim \wo^{-1}$ and hence the dimensionless indentation force $F\sim{\cal F}\wo^2\sim \wo$.

However, to make further progress we need to understand the behaviour of the solution in this region close to the outer region of its validity (i.e.~close to the inner radius of the wrinkle pattern, $\rho=\Li$, $\xi=1$). We note that \eqref{fvk1:tensileapprox} can be rearranged to give
\beq
\cF=\chi\frac{\upd\omega}{\upd\xi}=2\Phi\frac{\upd \Omega}{\upd\eta}=f\frac{\upd \Omega}{\upd\Phi}\left(1+A\Phi\right)^{1/2}\Phi^{1/2}
\eeq from which we find
\beq
\Omega=-1+\frac{2\cF}{A^{1/2}f}\sinh^{-1}(A\Phi)^{1/2}.
\eeq Crucially, we see that the rescaled vertical deflection at the inner portion of the wrinkled region $\Omega(\Phi=1)=-1+O(1/\wo)$: the boundary condition for the wrinkled problem at $\rho=\Li$ should, at leading order,  be simply $\wr=-\wo$ --- the imposed indentation depth. This is the result used in the simplified calculation presented in the main text.

As well as matching vertical displacements between regions, we must also match the slopes of the displacements (as a simple force balance argument illustrates). We therefore also need to know the value of $\wr'$ at $\rho=\Li^-$, which we find from \eqref{fvk1:tensileapprox} to be
\beq
\left. \frac{\upd \wr}{\upd \rho}\right|_{\rho=\Li^-}=\cF\wo^{3/2}/\lamI=\frac{F}{2\pi\mu\wo^{1/2}}=\frac{F}{2\pi C}.
\eeq

\subsection{The wrinkled region: $\Li<\rho<\Lo$ \label{sec:wrinkzone}}

In the wrinkled region we have $\Psi=C$ and that the vertical displacement satisfies \eqref{eqn:WrinkledFB}, which may be integrated immediately to give
\beq
\wr=\rho^2/2-\frac{\rho^3}{6C}+A+B\rho.
\label{eqn:wrinkprof1}
\eeq The constants $A$ and $B$ can be determined  from the continuity conditions at the inner boundary, $\rho=\Li$, where we have
\beq
\wr(\Li)=-\wo,\quad \left.\frac{\upd \wr}{\upd\rho}\right|_{\rho=\Li}=\frac{F}{2\pi C}.
\eeq Substituting these conditions into \eqref{eqn:wrinkprof1} and noting that $\Li\to0$ as $\wo\to\infty$ immediately gives $A=-\wo$, $B=F/(2\pi C)$ (the latter finding being consistent with the appearance of a Dirac $\delta$-function in \eqref{eqn:WrinkledFB}). Hence, we have
\beq
\wr=-\wo+\frac{F}{2\pi C}\rho+\rho^2/2-\frac{\rho^3}{6C}.
\label{eqn:wrinkprof2}
\eeq

We shall proceed to patch this profile onto that outside the wrinkled region, $\rho>\Lo$. In this regard, it is useful to note that differentiating the profile \eqref{eqn:wrinkprof2} and evaluating at $\rho=\Lo$ gives
\beq
\left. \frac{\upd \wr}{\upd \rho}\right|_{\rho=\Lo}=\frac{F}{2\pi C}+\Lo-\frac{\Lo^2}{2C}.
\eeq

\subsection{Beyond the wrinkled region: $\rho>\Lo$}

Eqn \eqref{fvk1:mem} can be integrated once to give
\beq
\frac{\upd \wr}{\upd\rho}=\rho-\frac{A+\rho^2/2}{\Psi}.
\eeq The constant of integration $A$ is immediately found by continuity of $\wr'$ and $\Psi$ at $\rho=\Lo$ to be $A=-F/2\pi$. (Again, this choice of the constant of integration is reassuring since it implies that the force being supported in the far field is that due to indentation, $F$.) We therefore have
\beq
\frac{\upd \wr}{\upd\rho}=\rho+\frac{F/\pi-\rho^2}{2\Psi}.
\label{eqn:outerslope}
\eeq

To solve the problem completely, however, we must determine the Airy stress function $\Psi$. To do this, we eliminate $\wr'$ from the equation of compatibility \eqref{fvk2:mem} using \eqref{eqn:outerslope} to give
\beq
\rho\frac{\upd}{\upd\rho}\left[\frac{1}{\rho}\frac{\upd}{\upd\rho}(\rho\Psi)\right]=\frac{1}{2}\left[\rho^2-\left(\frac{F/\pi-\rho^2}{2\Psi}\right)^2\right].
\label{eqn:outerPsieqn}
\eeq The solution of eqn \eqref{eqn:outerPsieqn} must match continuously onto that in the wrinkled region and so we must have that
\beq
\Psi(\Lo^+)=C,\quad\Psi'(\Lo^+)=0.
\eeq (Note that far away from the wrinkled region, we expect that $\Psi\sim\rho/2$, which will emerge necessarily from the structure of the equation.)

To make progress, we again make the substitution used in the inner tensile region, i.e.~$\Phi=\rho\Psi$, $\eta=\rho^2$, with which \eqref{eqn:outerPsieqn} becomes
\beq
\frac{\upd^2\Phi}{\upd\eta^2}=\frac{1}{8}\left[1-\left(\frac{F/\pi-\eta}{2\Phi}\right)^2\right],
\label{eqn:outerPhieqn}
\eeq with boundary conditions $\Phi(\lamO^2\wo)=\lamO\mu\wo$, $\Phi'(\lamO^2\wo)=\mu/(2\lamO)$ where $\lamO=\Li/\wo^{1/2}$. The condition at $\infty$ becomes that $\Phi\sim \eta/2$ as $\eta\to\infty$.

We rescale \eqref{eqn:outerPhieqn} by letting $\chi=\Phi/\wo$, $\xi=\eta/(\wo\lamO^2)$ so that
\beq
\wo^{-1}\frac{\upd^2\chi}{\upd\xi^2}=\frac{\lamO^4}{8}\left[1-\left(\frac{F/\pi\wo-\lamO^2\xi}{2\chi}\right)^2\right],
\label{eqn:outerChieqn}
\eeq with boundary conditions 
\beq
\chi(1)=\lamO\mu, \quad\chi'(1)=\mu\lamO/2,
\eeq and the far-field condition that $\chi\sim \lamO^2\xi/2$ as $\xi\to\infty$.

We note that
\beq
\chi=(\lamO^2\xi-F/\pi\wo)/2
\label{eqn:chiOuter}
\eeq is an exact solution of \eqref{eqn:outerChieqn} and, further, satisfies the  far-field condition, $\chi\sim\lamO^2\xi/2$,  as $\xi\to\infty$.
 However, this solution is incompatible with the boundary condition on $\chi'$ at $\xi=1$, i.e.~$\rho=\Lo$. We conclude that \eqref{eqn:chiOuter} is an outer solution with an inner, boundary layer-type solution close to $\xi=1$, $\rho=\Lo$.

\subsubsection*{The inner solution}

The outer solution \eqref{eqn:chiOuter} is incompatible with the boundary condition on $\chi'(1)$; we therefore look for a solution that is a small perturbation to the outer solution, but whose derivatives can be large enough to match this inner boundary condition. We therefore let
\beq
\chi=(\lamO^2\xi-F/\pi\wo)/2+\dchi,\quad \xi=1+X/\wo^{1/2},
\eeq  finding that
\beq
\frac{\upd^2\dchi}{\upd X^2}=\frac{\lamO^4}{2(\lamO^2-F/\pi\wo)}\dchi
\eeq and hence that
\beq
\chi=\tfrac{1}{2}(\lamO^2\xi-F/\pi\wo)+\alpha\exp\left[-\frac{\lamO^2}{2^{1/2}(\lamO^2-F/\pi\wo)^{1/2}}\left\{X=\wo^{1/2}(\xi-1)\right\}\right].
\eeq

The boundary condition on $\chi'(1)$ gives that
\beq
\alpha=\wo^{-1/2}\frac{\left(\lamO^2-F/\pi\wo\right)^{1/2}}{2^{3/2}\lamO}(\lamO+F/\pi\wo),
\eeq i.e. $\alpha=O(\wo^{-1/2})$. The boundary condition on $\chi(1)$ then gives, at leading order in $\wo^{-1/2}$, that
\beq
\mu=\lamO/2-\frac{F}{2\pi\lamO\wo}.
\label{eqn:mu}
\eeq

The key quantity of interest is actually $\wr'(\Lo)$, which we find, from $\Psi(\Lo)=\mu\wo^{1/2}$ and 
\beq
\left.\frac{\upd \wr}{\upd\rho}\right|_{\rho=\Lo+}=\Lo+\Lo\frac{F/\pi-\Lo^2}{(\Lo^2 -\frac{F}{\pi})}=0
\label{eqn:OuterInnerBC}
\eeq to leading order in $\wo^{-1/2}$.

With the remaining boundary condition at $\rho=\Lo^-$ determined by \eqref{eqn:OuterInnerBC} and continuity of slope, we now return to the wrinkled region to solve the remainder of the problem.

\subsection{Return to the wrinkled region}

In \S\ref{sec:largeD}\ref{sec:wrinkzone}, we  showed that the shape within the wrinkled region was given by
\beq
\wr=-\wo+\frac{F}{2\pi C}\rho+\rho^2/2-\frac{\rho^3}{6C}.
\eeq By considering the outer tensile region, we have now shown that
\beq
C/\wo^{1/2}=\mu=\lamO/2-\frac{F}{2\pi\lamO\wo}
\label{eqn:Candmu}
\eeq and that 
\beq
\wr'(\Lo^-)=0
\eeq to leading order in $\wo^{-1}$. Now, since, by the nature of the boundary layer, the derivative of a function is larger than the function itself in the boundary layer, we must also have $\wr(\Lo)=0$ to leading order, and hence that
\beq
\frac{F}{2\pi \wo}=\frac{\mu}{\lamO}\left[\frac{\lamO^3}{6\mu}-\lamO^2/2+1\right]=\frac{\lamO^2}{6}+\frac{\mu}{\lamO}(1-\lamO^2/2)
\eeq while \eqref{eqn:Candmu} gives us
\beq
\frac{F}{2\pi\wo}=\lamO\frac{\lamO-2\mu}{2}
\label{eqn:F}
\eeq and hence
\beq
\mu=\frac{2\lamO^3}{3(2+\lamO^2)}.
\label{eqn:mu}
\eeq

We also recall that from the inner tensile region we  had
\beq
\frac{\wo^2\cF^2}{\lamI\mu}=\frac{\lamI \left[F/(2\pi\wo)\right]^2}{\mu^3} = f^2\approx1.81527^2.
\label{eqn:lamI}
\eeq We have therefore determined relationships for the spring stiffness, $F/\wo$, $\lamI$ and $\mu$ all  in terms of $\lamO$. To close the problem we require an additional condition, which comes from considering the displacement across the wrinkled region.

\subsection{The final condition}

We recall the relationships for stress and strain in an axisymmetric shell are given by \cite[see][for example]{ventsel01}
\beq
\epsilon_{rr}=\frac{\pd u_r}{\pd r}+\frac{1}{2}\left(\frac{\pd \zeta}{\pd r}\right)^2+\frac{\zeta}{R}=\frac{\srr-\nu\sqq}{Y}
\label{eqn:epsrr}
\eeq and
\beq
\epsilon_{\theta\theta}=\frac{u_r}{r}+\frac{1}{2r^2}\left(\frac{\pd \zeta}{\pd \theta}\right)^2+\frac{\zeta}{R}=\frac{\sqq-\nu\srr}{Y}.
\label{eqn:epsqq}
\eeq The dimensionless versions of these equations give results for the horizontal displacement.

In the wrinkled region we use the condition on $\epsilon_{rr}$ to see that
\beq
\frac{\pd U_r}{\pd \rho}=\frac{C}{\rho}-\frac{1}{2}\left(\frac{\pd \wr}{\pd \rho}\right)^2-\wr
\label{eqn:Uwrink}
\eeq  while in the tensile (unwrinkled) regions it is simpler to use that from $\epsilon_{\theta\theta}$, in particular:
\beq
\frac{U_r}{\rho}=\Psi'-\nu\Psi/\rho-\wr.
\label{eqn:Utensile}
\eeq While these expressions hold in different regions, they must agree with the corresponding displacement within the wrinkled region, obtained by integrating \eqref{eqn:Uwrink}. Now, noting that $\Psi'(\Li^-)=\Psi'(\Lo^+)=0$ and $\Psi(\Li^-)=\Psi(\Lo^+)=C$, eqn \eqref{eqn:Utensile} yields:
\beq
U_r(\Li^-)+\Li Z(\Li^-)=U_r(\Lo^+)+\Lo Z(\Lo^+)=-\nu C.
\eeq It is therefore convenient to express the continuity of displacement in terms of the \emph{change in displacement} across the wrinkled region, i.e.
\beq
\Li Z(\Li^-)-\Lo Z(\Lo^-)=[U_r]_{\Li}^{\Lo}=\int_{\Li}^{\Lo}U_r'~\upd \rho.
\eeq We then find that
\beq
\Li \wr(\Li)-\Lo \wr(\Lo)=[U_r]_{\Li}^{\Lo}=C\log(\Lo/\Li)-\int_{\Li}^{\Lo}\frac{1}{2}\left(\frac{\pd \wr}{\pd \rho}\right)^2+\wr~\upd\rho
\eeq and hence, after integrating the last term in the integral on the RHS by parts:
\beq
C\log(\Lo/\Li)=\int_{\Li}^{\Lo}\frac{1}{2}\left(\frac{\pd \wr}{\pd \rho}\right)^2-\rho\frac{\pd \wr}{\pd \rho}~\upd\rho.
\label{eqn:dispcondition}
\eeq 

The condition \eqref{eqn:dispcondition} is the final condition required to close the system. It is implemented exactly in the numerical results  described in the main text. However, for the purposes of the theoretical analysis we  present here, it is enough to consider the scaling behaviour of the different terms. We find that the LHS of \eqref{eqn:dispcondition} $\sim \wo^{1/2}\log\wo $ while the RHS $\sim \wo^{3/2}$; we therefore conclude that at leading order
\beq
\int_{\Li}^{\Lo}\frac{1}{2}\left(\frac{\pd \wr}{\pd \rho}\right)^2-\rho \frac{\pd \wr}{\pd \rho}~\upd\rho=0.
\label{eqn:DispCancel}
\eeq Letting
\beq
\xi=\rho/\wo^{1/2},\quad \omega(\xi)=\wr/\wo
\eeq we have that $$\omega(\xi)=-1+\frac{F}{2\pi\mu\wo}\xi+\xi^2/2-\xi^3/(6\mu)$$ or
\beq
\frac{\wr_0(\rho)}{\wo}=-1+\frac{F}{2\pi\mu\wo^{1/2}}\frac{\rho}{\wo^{1/2}}+\tfrac{1}{2}\left(\frac{\rho}{\wo^{1/2}}\right)^2-\frac{1}{6\mu}\left(\frac{\rho}{\wo^{1/2}}\right)^3.
\label{eqn:Z0Prof}
\eeq
 Performing the integration over $\xi\in(0,\lamO)$ and making use of \eqref{eqn:F}-\eqref{eqn:mu} we find that
\beq
\lamO^4+24\lamO^2-36=0
\eeq i.e.
\beq
\lamO=6(\sqrt{5}-2)\approx1.19013,\quad \mu=\frac{2\lamO^3}{3(2+\lamO^2)}\approx0.328944,\quad \frac{F}{\wo}\approx1.990001.
\eeq

To calculate the value of $\lamI$ we then return to \eqref{eqn:lamI} from which we find that
\beq
\lamI\approx1.16923
\eeq

\subsection{The wrinkle amplitude \label{sec:slaving}}

We make an  ansatz for the perturbations of the shell about the tension-field state described above:
\beq
\wr(\rho,\theta)=\wr_0(\rho)+\tf(\rho)\cos m\theta
\label{eqn:WrinkAnsatz}
\eeq where $\tf(\rho)=f(\rho)\cdot R/\lp^2$.  In the determination of the energy, the form of the wrinkle amplitude, $f(r)$, will be of some interest and so we study this here, making use of the results for the mean shell displacement $\wr_0(\rho)$, \eqref{eqn:Z0Prof}.

Substituting the ansatz \eqref{eqn:WrinkAnsatz} into the stress-strain equations \eqref{eqn:epsrr}--\eqref{eqn:epsqq} we find that
\beq
\frac{\pd U_r}{\pd \rho}=\frac{C}{\rho}-\frac{1}{2}\left(\frac{\upd \wr_0}{\upd \rho}\right)^2-\wr_0
\label{eqn:dudrho}
\eeq and also the ``slaving condition"
\beq
\frac{m^2}{4}\frac{\tf^2}{\rho^2}=-\nu\frac{C}{\rho}-\wr_0-\frac{U_r}{ \rho}.
\label{eqn:uslave}
\eeq Eqn \eqref{eqn:dudrho} may be integrated once, subject to the condition that $U_r(\Lo)=-\nu C-\wr_0(\Lo)\Lo$,  to determine $U_r(\rho)$ within the wrinkled region. $U_r$ may then be eliminated from \eqref{eqn:uslave} to give
\begin{align}
\frac{m^2}{4}\frac{\tf^2}{\rho^2}&=-\wr_0+\frac{1}{ \rho}C\log\frac{\Lo}{\rho}-\frac{1}{\rho}\int_\rho^{\Lo}\left[\wr_0+\tfrac{1}{2}\left(\frac{\upd\wr_0}{\upd\rho}\right)^2\right]~\upd\rho\nonumber\\
&=\frac{1}{ \rho}C\log\frac{\Lo}{\rho}+\frac{1}{\rho}\int_\rho^{\Lo}\left[\rho\wr_0'-\tfrac{1}{2}\left(\frac{\upd\wr_0}{\upd\rho}\right)^2\right]~\upd\rho
\end{align} (where we have used $\wr_0(\Lo)=0$ to eliminate the constant of integration).

The integration is readily performed, giving
\beq
m^2\tf^2=\left(36-84/\sqrt{5}\right)\wo~\xi\log\xi+9(7-3\sqrt{5})\wo^2~\xi^2(1-\xi^2)^2,\quad \xi=\rho/\Lo.
\label{eqn:Slaving}
\eeq

\subsection{Summary\label{sec:summary}}

In summary we have the asymptotic predictions for the wrinkle positions
\beq
\Li\approx1.16923\wo^{-1/2}
\label{eqn:FTin}
\eeq and
\beq
\Lo\approx 1.19013\wo^{1/2}
\label{eqn:FTout}
\eeq while the predicted force law is
\beq
F\approx 1.99 \wo.
\label{eqn:FTforce}
\eeq

The mean shape of the shell (underlying the wrinkles) in the limit of large indentations may be written
\beq
\frac{\wr_0(\rho)}{\wo}=-1+\tfrac{1}{4}\left(6-\lambda^2\right)\xi+\tfrac{1}{2}\lambda^2\xi^2-\tfrac{1}{4}(2+\lambda^2)\xi^3\approx-1+ 1.146\xi+0.708\xi^2-0.854\xi^3
\label{eqn:MeanShape}
\eeq where $\xi=\rho/\Lo=\rho/(\lambda\wo^{1/2})$. Equation \eqref{eqn:MeanShape} is given as (9) in the main text. Finally, the slaving condition \eqref{eqn:Slaving} in this limit becomes
\beq
m^2\tf^2\approx9(7-3\sqrt{5})\wo^2\xi^2(1-\xi^2)^2.
\label{eqn:AsyIsoSlaving}
\eeq

\section{The sub-dominant energy \label{sec:subdom}}

To estimate the number of wrinkles observed, and to understand better the emergence of `asymptotic isometry' in this problem, we consider the energy associated with the wrinkle pattern. Unlike the energies $\Uelast$ and $\Ugas$ discussed in the main text, this energy vanishes in the limit of vanishing thickness and so is referred to as the `sub-dominant' energy \cite[][]{davidovitch11}. More precisely, in contrast to those ``dominant" energies (discussed in the main text), the sub-dominant energies are proportional to the (square of) wrinkle amplitude $f$. In principle, there are three separate contributions to the  sub-dominant energy due to  wrinkly undulations: elastic strain (stretching, beyond the energy $\Uelast$ that is discussed in the main text, that is only associated with the strain in the radial profile), bending energy and the excess work done in compressing the gas (beyond $\Ugas$ discussed in the main text). In practice, the latter energy is proportional to the volume of the deformation and so, assuming a sinusoidal wrinkle pattern, there is no additional energy stored in the gas as a result of wrinkling. Here, we focus on the former two sources of sub-dominant energy.

In this section we use dimensional variables for convenience.

\subsection{Sub-dominant stretching energy}

The requirement that the shear-stress and strain vanish to leading order, $\srq=\epsilon_{r\theta}=0$, shows that within the wrinkled region the correction to the radial stress, $\srro=\srr-C/r$ is linear in the wrinkle amplitude $f(r)$ with
\beq
\srr= pR\lp\cdot\frac{C}{r}+Yf\left(\frac{1}{R}-\frac{\upd^2\zeta_0}{\upd r^2}\right)\cos m\theta,
\eeq while the radial displacement may be written
\beq
u_r=u_r^{(0)}-f\zeta_0'\cos m\theta.
\eeq
 This result generalizes an earlier result for a naturally flat sheet adhered to a stiff spherical substrate \cite[][]{hohlfeld15} to the case of a naturally curved shell.

To calculate the elastic energy of the system, we must integrate the strain energy density, $\srr\epsilon_{rr}/2$ over both $r$ and $\theta$. We are therefore interested only in the parts that will persist upon integrating wrt $\theta$ (i.e.~those that involve even powers of $\cos m\theta$). We have that
\begin{align}
\epsilon_{rr}&=\frac{\partial u_r}{\partial r}+\tfrac{1}{2}\left(\frac{\partial \zeta}{\partial r}\right)^2+\frac{\zeta}{R}\nonumber\\
&=\epsilon_{rr}^{(0)}+\tfrac{1}{2}\bigl[f'\cos m\theta\bigr]^2+\left(\frac{1}{R}-\frac{\upd^2\zeta_0}{\upd r^2}\right)f\cos m\theta
\end{align}

The azimuthally averaged energy density is therefore 
\beq
\overline{\srr\epsilon_{rr}}=pR\lp \frac{C}{r}\epsilon_{rr}^{(0)}+\tfrac{1}{4}(f')^2 \frac{pR\lp \cdot C}{r}+\tfrac{1}{2}Y f^2\left(\frac{1}{R}-\frac{\upd^2\zeta_0}{\upd r^2}\right)^2.
\label{eqn:radialenergy}
\eeq 

In the language of the FT approach to wrinkling \cite[][]{davidovitch11},  the dominant energy is the first of the three terms in \eqref{eqn:radialenergy} with the other two being the contributions to the sub-dominant energy, i.e.~the energy associated with wrinkles themselves, which vanish with the wrinkles' amplitude $f$. The first of these sub-dominant terms corresponds to the radial stretching of wrinkles, the second to the curvature of the wrinkled state relative to its natural configuration.

We use the dimensionless versions of $f$ and $\zeta_0$ to write the two components of the strain energy as
\beq
\overline{\srr\epsilon_{rr}}^{(\mathrm{sub})}=\frac{p^2R^2}{Y}\left[ \tfrac{1}{4}(\tf')^2 \frac{\mu\wo^{1/2}}{\rho}+\tfrac{1}{2} \tf^2\left(1-\frac{\upd^2\wr}{\upd \rho^2}\right)^2\right].
\label{eqn:radialenergydimless}
\eeq 

We then have that the stretching energy is given by
\begin{eqnarray}
U_s&=&\pi\int_{\Li}^{\Lo}\overline{\srr\epsilon_{rr}}~r~\upd r\nonumber\\
&=&\pi\frac{p^3R^5}{Y^2}\Lo^2\int_{\Li/\Lo}^{1}\left\{\tfrac{1}{4}\wo(\tf_\xi)^2 \frac{\mu}{\lambda}+\tfrac{1}{2} \tf^2\left(1-\lambda^{-2}\frac{\upd^2\wr}{\upd \xi^2}\right)^2\right\} \xi~\upd \xi\label{eqn:Ustretch2}\\
&\propto& \frac{p^3R^5}{Y^2}\wo^3m^{-2} \label{eqn:UstretchFinal},
\end{eqnarray} where we have used the slaving condition,  \eqref{eqn:AsyIsoSlaving}.  We note that this energy is dominated by the second term on the RHS of \eqref{eqn:Ustretch2}.

\subsection{Bending energy}

The bending energy may be written using the slaving condition for very large indentation depths, \eqref{eqn:AsyIsoSlaving}. We find:
\begin{eqnarray}
U_b&=&\tfrac{1}{2}B\int_0^{2\pi}\int_{\Li}^{\Lo}r(\nabla^2\zeta)^2~\upd r~\upd \theta\approx\tfrac{\pi}{2}B\int_{\Li}^{\Lo}r\bigl[m^2f(r)/r^2\bigr]^2~\upd r\nonumber\\
&=&m^2\frac{\pi B}{2} \frac{\lp^2}{R^2}\Lo^{-2}\int_{\Li/\Lo}^{1}\xi^{-3}\bigl[m \tf(\xi)\bigr]^2~\upd \xi\nonumber\\
&\propto& \frac{B pR}{Y}m^2\wo\log\wo
\end{eqnarray} at leading order in $\wo$.

\subsection{Predicted wrinkle number}

The sum of elastic energies, $\Utot=U_s+U_b$ is a function of the number of wrinkles, $m$. This sum is minimized by
\beq
m^4\propto\frac{p^2R^4}{YB}\frac{\wo^2}{\log\wo}=\frac{\tau^2\wo^2}{\log\wo}
\eeq and is precisely equation (18) of the main text.

\section{Comparison of floating films to pressurized shells \label{sec:compare}}

Some of the behaviour exhibited in the wrinkling of a pressurized elastic shell is similar to what was seen in a recent study of the indentation of a thin elastic film floating on a liquid bath and indented at a point \cite[][]{vella15}. In this section we discuss the similarities and differences between the two scenarios.

Details of the experiment are given by \cite{vella15}. In brief, thin Polystyrene films of Young modulus $E=3.4\mathrm{~GPa}$, Poisson ratio $\nu=0.3$, radius $\Rfilm=1.14\mathrm{~cm}$ and thickness in the range $85\mathrm{~nm}\leq t\leq 246\mathrm{~nm}$ were floated at the surface of deionized water, with density $\rhol$. In the floating state, the surface tension of the water, $\glv$, applies a tension at the boundary of the film and so the stress is initially uniform and isotropic, $\sinf=\glv$. The freely floating films were then indented at a point with wrinkling beginning at a threshold indentation depth $\wo_c\sim \lc (\glv/Et)^{1/2}$ where $\lc=(\glv/\rhol g)^{1/2}$ is the capillary length of the bare liquid--gas interface. As Table \ref{table:compare} shows, this behaviour is very similar to what is observed in the pressurized shell example with the modifications that in this case the foundation stiffness $\kfound=\rhol g$ arises from hydrostatic pressure, and the background tension $\sinf=\glv$, arises from the surface tension of the liquid.

In the floating film case, the outer edge of the wrinkles very quickly reach the edge of the sheet, $\Lo=\Rfilm$, at which point the system begins to be influenced by a wrinkly isometry similar to that observed here. However, since the outer edge of the wrinkles is fixed, the length scale over which the film returns to be flat, $\lcurv=(\Rfilm^2\lc)^{1/3}$, is independent of indentation depth. A key qualitative difference between the Wrinkly Isometry observed in the floating film and the pressurized shell is that in the former there is no unwrinkled isometry (except a non-axisymmetric ``developable cone"  involving delamination of the sheet from the liquid sub-phase); in the latter, mirror buckling remains as an allowed, unwrinkled, isometry (though, as seen here, it has a higher energetic cost than Wrinkly Isometry).

One interesting feature of the shell problem is that the radial displacement across the sheet is not, in fact $\delta^2/\lcurv$, as was found in the floating sheet problem, and might be expected intuitively \cite[see][for example]{vella15}. The difference between the two problems can be understood as follows: In the floating film problem, wrinkles extend to the edge of the sheet and there is no tensile region \emph{beyond} the wrinkles. Because of this, the displacement must be determined by integrating the strain in the wrinkled region, which gives a purely geometric result (i.e.~one that does not depend explicitly on the exerted tension), as expected on the basis of asymptotic isometry. For the shell case, there is always an outer tensile region. As such, the strain relationships immediately give that $u_r(\Lo)\sim C\sim \delta^{1/2}$. The difference between the two problems could be attributed to the fact that  a naturally curved object can accommodate deflection without horizontal displacement because of its curvature (just think of the mirror buckling solution in which points merely get reflected and so do not move horizontally, even though there is a vertical deflection).

\begin{table}
\begin{center}
\begin{tabular}{|c|cc|c|}
    \hline
   \multirow{2}{*}{\textbf{Property}}&\multicolumn{2}{c|}{\textbf{Pressurized Shell}} & \textbf{Floating Membrane}\\
  & MB & WI & WI\\
    \hline
    $\sinf$ & \multicolumn{2}{c|}{$pR/2$}  & $\glv$\\
    \hline
\specialcell{Foundation stiffness\\ $\kfound$}    & \multicolumn{2}{c|}{$\sim Et/R^2$} & $\rhol g$\\
\hline
\specialcell{Capillary length\\ $\bigl[\sim(\sinf/\kfound)^{1/2}\bigr]$} &  \multicolumn{2}{c|}{$\lp$} & $\lc=(\glv/\rhol g)^{1/2}$ \\
\hline
\specialcell{Wrinkle onset \\ $\bigl[\wo_c\sim \lc (\sinf/Et)^{1/2}\bigr]$} & \multicolumn{2}{c|}{$\lp^2/R$} & $\lc (\glv/Y)^{1/2}$ \\
\hline
         $\Lo$ & $=(\delta R)^{1/2}$ & $\approx1.19(\delta R)^{1/2}$ & $\Rfilm$\\
         \hline
	$\lcurv$ & $\Lo$ & $\Lo$ & $\ell_\ast=(\Rfilm^2\lc)^{1/3}$\\
             \hline
     $\Li$ & increasing with $\delta$ & $\sim \delta^{-1/2}$ & $\sim\delta^{-2}$ \\ 
     \hline
  \specialcell{Indentation stiffness\\ $\bigl[k=F/\delta\bigr]$}  & $=\pi pR$ & $\approx 1.99pR$ & $\sim (\glv^2\rhol g \Rfilm^2)^{1/3}$\\
    \hline

    $u(\Lo)-u(\Li)$ & $\sim \lp^2\delta^{1/2}/R^{3/2}$ & $\sim \lp^2\delta^{1/2}/R^{3/2}$ & $\delta^2/\lcurv\sim\delta^2/\ell_\ast$\\
    \hline
\end{tabular}
\end{center}
\caption{Comparison between the two types of isometry of a pressurized shell, i.e.~Mirror Buckling (MB) and Wrinkly Isometry (WI), and also the Wrinkly Isometry of a floating elastic membrane \cite[studied in detail by][]{vella15}.}
\label{table:compare}
\end{table}